\documentclass[11pt,a4paper,onecolumn,numbers]{elsarticle}

\usepackage[latin1]{inputenc}
\usepackage{amsmath}
\usepackage{amsfonts}
\usepackage{amssymb}
\usepackage{graphicx}
\usepackage{fancyhdr}
\usepackage{times}
\usepackage{natbib}

\begin{document}
\begin{frontmatter}

\title{Investigation on the energy and mass composition of cosmic
rays using LOPES radio data}


\begin{keyword}
radio detection\sep  cosmic rays\sep  air showers\sep  LOPES \sep Xmax \sep primary energy
\PACS 96.50.sd\sep  95.55.Jz
\end{keyword}

\author[2]{N.~Palmieri\corref{cor}}
\ead{Nunzia.Palmieri@kit.edu}
\author[1]{W.D.~Apel}
\author[2,14]{J.C.~Arteaga}
\author[3]{L.~B\"ahren}
\author[1]{K.~Bekk}
\author[4]{M.~Bertaina}
\author[5]{P.L.~Biermann}
\author[1,2]{J.~Bl\"umer}
\author[1]{H.~Bozdog}
\author[6]{I.M.~Brancus}
\author[4]{A.~Chiavassa}
\author[1]{K.~Daumiller}
\author[2,15]{V.~de~Souza}
\author[4]{F.~Di~Pierro}
\author[1]{P.~Doll}
\author[1]{R.~Engel}
\author[3,9,5]{H.~Falcke}
\author[2]{B.~Fuchs}
\author[10]{D.~Fuhrmann}
\author[11]{H.~Gemmeke}
\author[7]{C.~Grupen}
\author[1]{A.~Haungs}
\author[1]{D.~Heck}
\author[3]{J.R.~H\"orandel}
\author[5]{A.~Horneffer}
\author[2]{D.~Huber}
\author[1]{T.~Huege}
\author[1,16]{P.G.~Isar}
\author[10]{K.H.~Kampert}
\author[2]{D.~Kang}
\author[11]{O.~Kr\"omer}
\author[3]{J.~Kuijpers}
\author[2]{K.~Link}
\author[12]{P.~{\L}uczak}
\author[2]{M.~Ludwig}
\author[1]{H.J.~Mathes}
\author[2]{M.~Melissas}
\author[8]{C.~Morello}
\author[2]{J.~Oehlschl\"ager}
\author[1]{T.~Pierog}
\author[10]{J.~Rautenberg}
\author[1]{H.~Rebel}
\author[1]{M.~Roth}
\author[11]{C.~R\"uhle}
\author[6]{A.~Saftoiu}
\author[1]{H.~Schieler}
\author[11]{A.~Schmid}
\author[1]{F.G.~Schr\"oder}
\author[13]{O.~Sima}
\author[6]{G.~Toma}
\author[8]{G.C.~Trinchero}
\author[1]{A.~Weindl}
\author[1]{J.~Wochele}
\author[1]{M.~Wommer}
\author[12]{J.~Zabierowski}
\author[5]{J.A.~Zensus}
\author{\\ - LOPES Collaboration -}

\address[1]{Institut f\"ur Kernphysik, Karlsruhe Institute of Technology (KIT), Germany}
\address[2]{Institut f\"ur Experimentelle Kernphysik, KIT - Karlsruher Institut f\"ur Technologie, Germany}
\address[3]{Radboud University Nijmegen, Department of Astrophysics, The Netherlands}
\address[4]{Dipartimento di Fisica Generale dell' Universit\`a di Torino, Italy}
\address[5]{Max-Planck-Institut f\"ur Radioastronomie Bonn, Germany}
\address[6]{National Institute of Physics and Nuclear Engineering, Bucharest, Romania}
\address[7]{Fachbereich Physik, Universit\"at Siegen, Germany}
\address[8]{Istituto di Fisica dello Spazio Interplanetario, INAF Torino, Italy}
\address[9]{ASTRON, Dwingeloo, The Netherlands}
\address[10]{Fachbereich Physik, Universit\"at Wuppertal, Germany}
\address[11]{Institut f\"ur Prozessdatenverarbeitung und Elektronik, KIT - Karlsruher Institut f\"ur Technologie, Germany}
\address[12]{Soltan Institute for Nuclear Studies, Lodz, Poland}
\address[13]{Department of Physics, University of Bucharest, Bucharest, Romania}
\address[14]{now at: Univ Michoacana, Morelia, Mexico}
\address[15]{now at: Univ S$\tilde{a}$o Paulo, Inst. de F\'{\i}sica de S\~ao Carlos, Brasil}
\address[16]{now at: Inst. Space Sciences, Bucharest, Romania }

\begin{abstract}

The sensitivity to the mass composition as well as the reconstruction of the energy of the primary particle are explored here by leveraging the features of the radio lateral distribution function.
For the purpose of this analysis, a set of events measured with the LOPES experiment is
reproduced with the latest CoREAS radio simulation code.
Based on simulation predictions, a method which exploits the slope of the radio lateral distribution function is developed (Slope Method) and directly applied on measurements.
As a result, the possibility to reconstruct both the energy and the X$_{\mathrm{max}}$, i.e. depth of the shower maximum, of the cosmic ray air shower using radio data and achieving relatively small uncertainties is presented.

\end{abstract}
\end{frontmatter}

\section{Introduction}
A precise reconstruction of both energy and mass composition of the primary cosmic rays still remains a key goal in astroparticle physics. In the recent past, the detection of the coherent radio emission from extensive air showers at MHz frequencies as well as the understanding of its emission mechanisms have reached impressive results.\\
 As predicted from pure simulations, the information about the depth of the shower maximum (i.e.  X$_{\mathrm{max}}$ )- and, thus, indirectly about the primary type - can be directly extracted from the slope of the radio lateral distribution function (LDF) \cite{REAS2,Krijn,LudwigHuege}. This dependence can be easily referred to a geometrical effect: iron nuclei interact earlier in the atmosphere and the showers develop faster compared to proton primaries; the radio source for an iron event is, thus, further away from an observer at ground, and this implies a flatter slope of the lateral distribution function compared to proton events.\\
 A method (Slope Method), which achieves information on the primary energy and the X$_{\mathrm{max}}$ of the air shower by analysing the features of the radio LDF, has previously been successfully developed and applied on older simulations (REAS3) \cite{PalmieriArena,PalmieriThesis}. More sophisticated and more complete CoREAS simulations of LOPES events are now available, which also include a realistic treatment of the refractive index of the atmosphere.  \\
 Taking the analysis in \cite{PalmieriArena,PalmieriThesis} as a guideline, the Slope Method based on CoREAS simulations of LOPES events is developed, further improved and applied to LOPES data. The main results are presented in the following.

\section{LOPES event selection}
LOPES \cite{lopes} is a digital interferometric radio antenna array co-located with the particle detector experiment KASCADE-Grande \cite{kascade} at KIT, Germany. 
The following analysis includes events measured with two LOPES setups, LOPES 30 and LOPES 30 pol. \cite{timArena}, which consisted, respectively, of 30 and 15 calibrated dipole antennas oriented in the east-west direction.\\
The co-location with KASCADE-Grande brings as a profit the reconstruction of fundamental air shower parameters, such as energy, core position and arrival direction, for each single event.
After quality cuts on the signal-to-noise ratio, on the coherence of the radio signal, and on the goodness of the fit for the radio lateral distribution function, over 200 events remains for the analysis. On average, the primary energy is of $\sim 10^{17}$eV, the zenith angle is less than 40 degrees, and the maximum shower core distance is 90 m.\\
 
For each LOPES measured event, one proton-generated air shower and one iron-generated air shower is produced with CoREAS \cite{CoREASArena} (QGSJet II.03 is used as hadronic interaction model). 
For the purpose of this analysis, the simulations are created in order to reproduce a realistic case: on the one hand, information about the LOPES selected events, such as core position, primary energy, and incoming direction reconstructed by KASCADE-Grande are used as input parameters for the CORSIKA showers \cite{corsikaLudwig}. On the other hand, shower-to-shower fluctuations are included, since there is no pre-selection of a typical shower, i.e. shower with a typical X$_{\mathrm{max}}$.\\ In order to represent best the recorded event, a further step is made in the pre-selection of the CORSIKA showers: the number of muons (N$\mu$) measured by KASCADE-Grande is used as discriminator for this purpose. 200 CONEX showers for proton and 100 for iron are simulated with QGSJet II.03 and UrQMD respectively for high and low energy interaction \cite{corsikaLudwig}. The CONEX shower which can best reproduce the measured N$\mu$ is chosen and simulated with CoREAS. In this way a specific shower similar to what has been detected by KASCADE-Grande is used.\\
Moreover, the observer positions for the CoREAS-simulated LDF represent the real arrangement of the LOPES antennas in the field with respect to the core of the shower.\\ 

 \begin{figure}[t!]
  \vspace{5mm}
  \centering
  \includegraphics[width=1\textwidth]{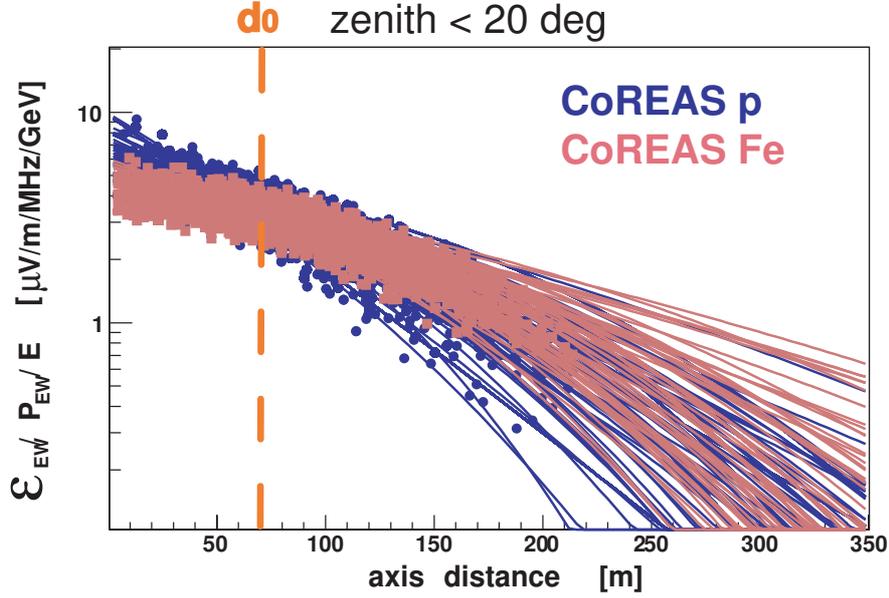}
  \caption{Normalized CoREAS lateral distribution function for the events with zenith angle smaller than 20 deg., simulated once as proton (blue), once as iron (light-red). The points are fitted with a Gaus-function.}
   \label{LDF}
\end{figure} 
\section{Slope Method on CoREAS}
CoREAS simulations are one of the last achievements of a better understanding and an improved modelling of the radio emission mechanisms from air showers \cite{CoREASArena}. With a realistic treatment of the refractive index of the atmosphere, CoREAS simulations are more complete than the predecessors and take into account the Cherenkov-like behaviour of the radio emission, which is the main reason for the flattening of the radio lateral distribution function close to the shower axis \cite{lopesLDF}. The flattening strongly depends on both the geometry of the air shower and on the primary type.
A comparison between CoREAS simulations and LOPES measurements is presented in \cite{SchroederLOPES_ICRC2013}.  \\
With the CoREAS simulations, i.e. the realistic treatment of the atmospheric refractive-index, the radio lateral distributions function (LDF) at the distances probed by the LOPES experiment becomes much more complex. For the purpose of this analysis, we use the Gauss-function \begin{equation}
\epsilon (d) \simeq \epsilon_{\mathrm{G}} \exp \left(\frac{(d - b)^{2}}{2c^{2}} \right)
   \label{eqLDF}
\end{equation}
with $\epsilon_{\mathrm{G}}$ [$\mu$V/m/MHz], $b$ [m], and $c$ [m] the 
free parameters to fit the LOPES LDF. Eq.\ref{eqLDF} describes the lateral distributions better than the simple (two-parameter) exponential function used in other LOPES analyses \cite{PalmieriArena,PalmieriThesis}. \\
The influence on the LDF slope of the zenith angle is taken into account by separately looking at 5 different zenith angle bins. The first, which includes zeniths between 0 and $\sim$20 degrees, is shown in this paper as example.\\   
This method aims to compare LDFs of events with different primary energies and arrival directions, thus normalizations for the radio amplitudes are required (fig.\ref{LDF}). The primary energy reconstructed by KASCADE-Grande is used for the energy normalization. With the reasonable assumption that the predominant contribution to the radio emission in air showers has geomagnetic origin \cite{Horneffer2007}, the arrival direction is taken into account by correcting for \textbf{P$\mathrm{_{EW}}$}, the east-west component of the vector $\textbf{v}\times\textbf{B}$, with $\textbf{v}$ velocity and $\textbf{B}$ the geomagnetic field vector.\\ 
\subsection{Primary Energy}
 \begin{figure}[t!]
  \vspace{5mm}
  \centering
  \includegraphics[width=1\textwidth, height=0.7\textwidth]{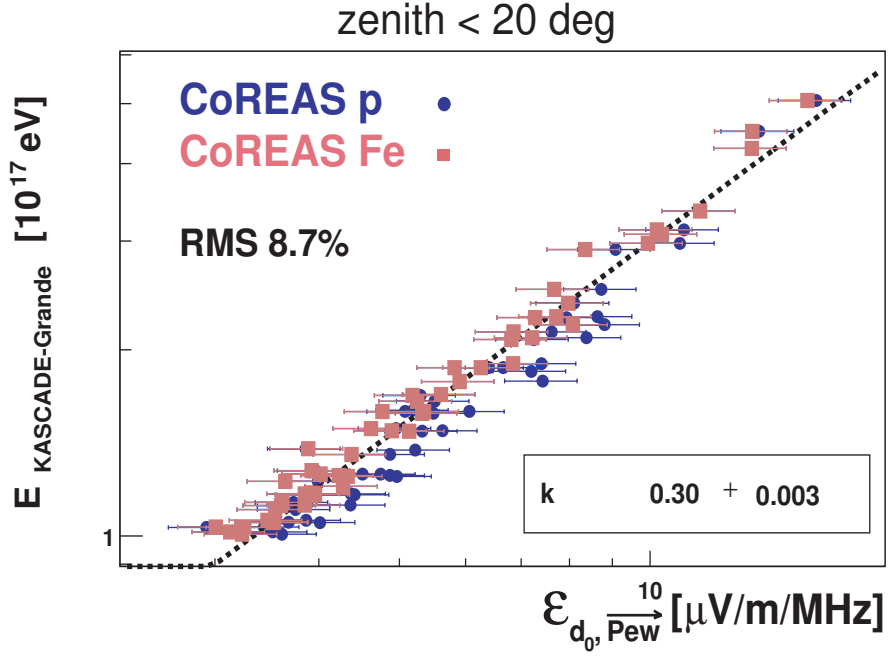}
  \caption{Linear correlation between the reconstructed primary energy and the CoREAS radio pulse in d$_{0}$, with k the free parameter of the fit. The radio amplitude is normalized by the east-west component of the $\textbf{v}\times\textbf{B}$ vector. The RMS-spread is $\sim8.7\%$.}
   \label{REAS_energy_lin}
\end{figure} 
Similar to our previous analysis \cite{PalmieriArena}, a specific distance from the shower axis (d$_{0}$) is identified where the LDF profiles exhibit the smallest RMS spread. In d$_{0}$ the radio amplitude is barely affected by shower-to-shower fluctuations and it does not carry any information about the primary type \cite{REAS2,PalmieriArena}. This region is identified by looking at the RMS spread of the LDF fits at several distances from the shower core and picking the distance with the smallest RMS value. For the LOPES events d$_{\mathrm{0}}$ lies in the distance range between 70 m and 100 m from the shower axis \footnote{in shower plane coordinate system} depending on the zenith angle bin. For the events with zenith smaller than 20 deg, d$_{\mathrm{0}}$ is marked by the dashed line in fig.\ref{LDF}, (70 m).\\
A direct correlation between the radio amplitude in d$_{\mathrm{0}}$ - normalized for the arrival direction - and the energy of the primary is expected \cite{REAS2,PalmieriArena}.\\
For the LOPES simulated events a linear correlation with the reconstructed primary energy from KASCADE-Grande is shown in fig.\ref{REAS_energy_lin}. The spread around the linear fit in fig.\ref{REAS_energy_lin} indicates the intrinsic uncertainty of the Slope Method in determining the primary energy. For the complete LOPES selection, including all zenith angle bins, an average RMS spread of $\sim9\%$ is found.
\subsection{X$_{\mathrm{max}}$ correlation}
 \begin{figure}[h!]
  \vspace{5mm}
  \centering
  \includegraphics[width=1\textwidth, height=0.7\textwidth]{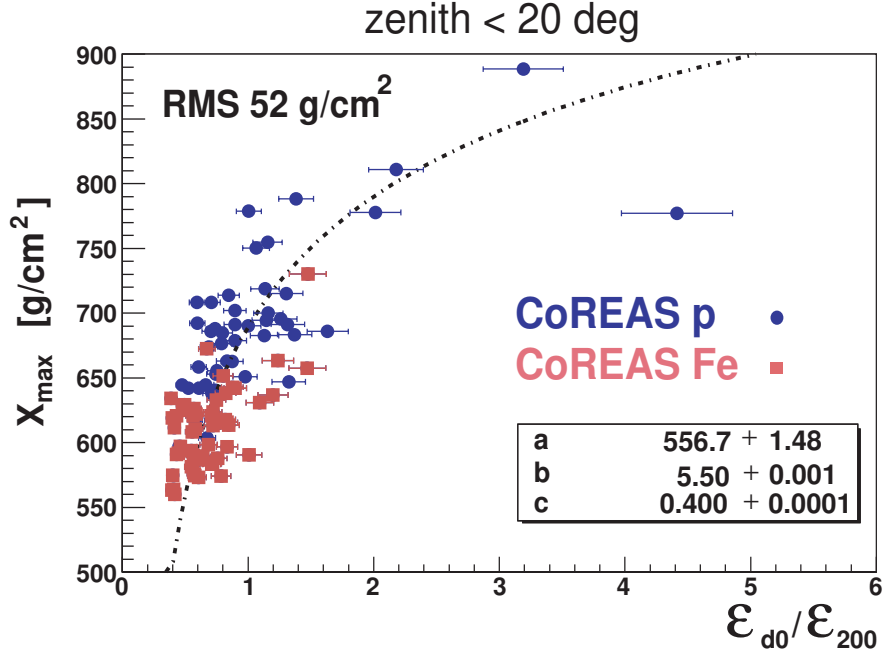}
  \caption{Correlation between the true X$_{\mathrm{max}}$ from CORSIKA simulations and the radio simulated LDF slope. The RMS-spread is of 52 ~g/cm$^{2}$.}
   \label{xmax_fit_1bin}
\end{figure} 
The slope of the radio LDF is the feature of the Gauss-fit which best correlates with the depth of the shower maximum (X$_{\mathrm{max}}$). The slope is defined as the ratio of the radio amplitudes at d$_{0}$ and at 200 m from the shower axis ($\frac{\epsilon_{d0}}{\epsilon_{d 200}}$). The dependence of the slope on the zenith angle is not clearly identified, therefore the usual five zenith angle bins are used. The correlation of fig.\ref{xmax_fit_1bin} has been fit with
\begin{equation}
X_{\mathrm{max}} = a \left[ \ln (b \frac{\epsilon_{d0}}{\epsilon_{200}}) \right]^{c}
\label{xmaxformula}
\end{equation}
already successful in previous analyses  \cite{REAS2,PalmieriArena}.\\
 For the complete zenith angle range an uncertainty of 50-70~g/cm$^{2}$ is predicted (RMS-spread of the fit), with the larger values due to the larger zenith angles.\\
In order to better constrain the fit and to improve the predicted uncertainty (~50~g/cm$^{2}$), an analysis which includes higher statistics for CoREAS simulations is planned. 
 \\
The values for the three fitting parameters ($a$, $b$ and $c$) will be used in the following to reconstruct X$_{\mathrm{max}}$ with the LOPES measurements.

\section{Slope Method on LOPES measurements}
 \begin{figure}[t!]
  \vspace{5mm}
  \centering
  \includegraphics[width=1.1\textwidth, height=0.7\textwidth]{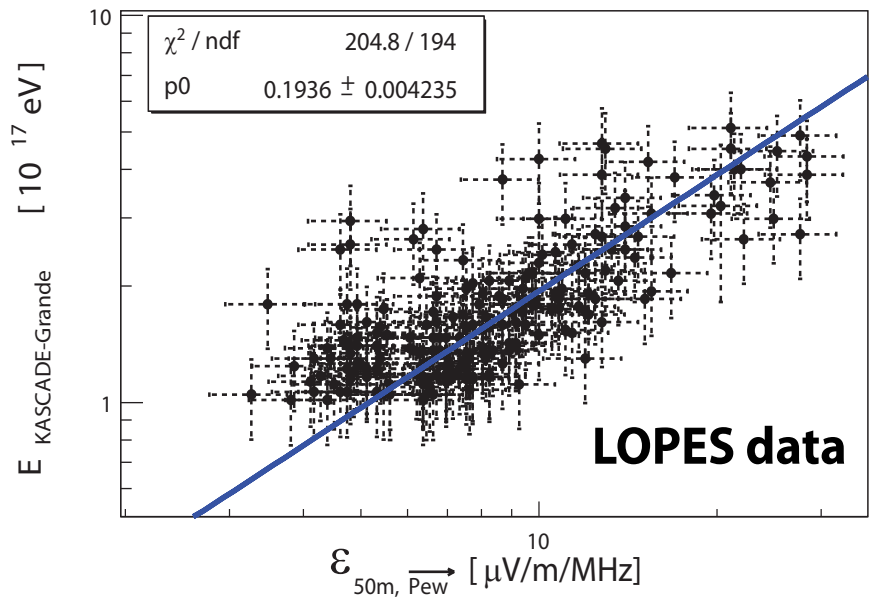}
  \includegraphics[width=1\textwidth]{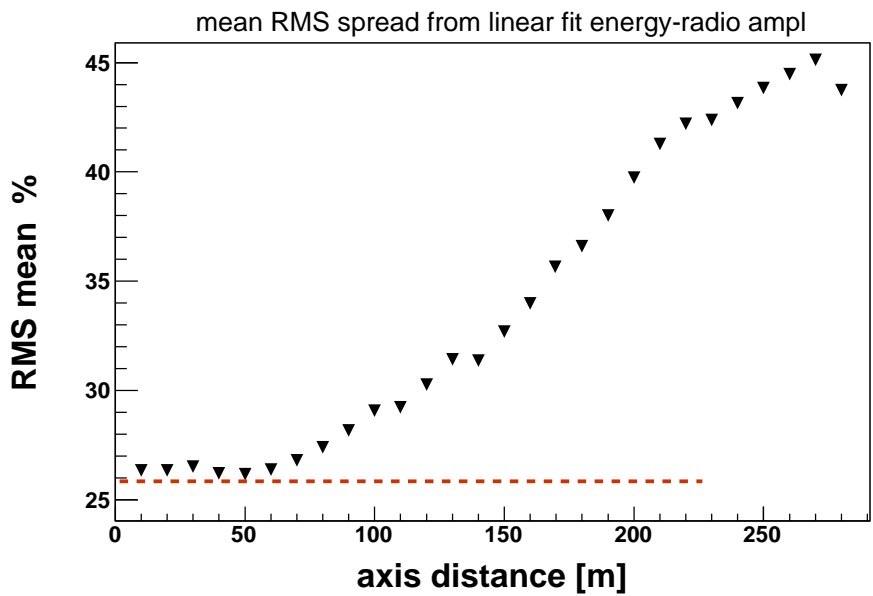}
  \caption{Top: Linear correlation of the KASCADE-Grande reconstructed primary energy and the LOPES measured radio pulse at 50 m, normalized for Pew. The RMS-spread if of 26\% Bottom: RMS-spread from the energy-fit averaged over the zenith bins, at several distances from the shower axis. With the measurements d$_{\mathrm{0}}$ is predicted around 50 m. }
   \label{figenergy}
\end{figure}
One important goal of this analysis is the confirmation in the LOPES data of d$_{\mathrm{0}}$ as the best distance from the shower axis for the radio amplitude--energy correlation. Indeed, at such distance, the highest precision on the energy reconstruction may be achieved.
Each LOPES LDF is fit with a Gauss function as well. The radio amplitude at a specific distance from the shower axis, derived from the fit value at such distance, is correlated with the primary energy reconstructed by KASCADE-Grande (in fig.\ref{figenergy}--top is reported as example the radio amplitude taken at 50~m from the shower axis). Several distances from the shower axis are considered, and for each the corresponding radio amplitude-energy linear correlation. The RMS-spread around each linear fit is again an indicator of the precision for the primary energy reconstruction achievable at each distance. The RMS-spreads obtained for these several distances, and averaged on the 5 zenith angle bins, are shown in fig.\ref{figenergy}--bottom. The minimum value, of about 26\%, is reached at distances less than 100 m, as expected from the CoREAS simulations, confirming the importance of d$_{\mathrm{0}}$ for the primary energy investigation. Moreover, the minimum spread of about 26\%, which contains the combination of the LOPES and of the KASCADE-Grande energy uncertainties, it is still comparable with the statistical uncertainty of KASCADE (20\%--40\%) \cite{kascade}.\\
 \begin{figure}
  \hspace{10cm}
  \centering
  \includegraphics[width=1.4\textwidth]{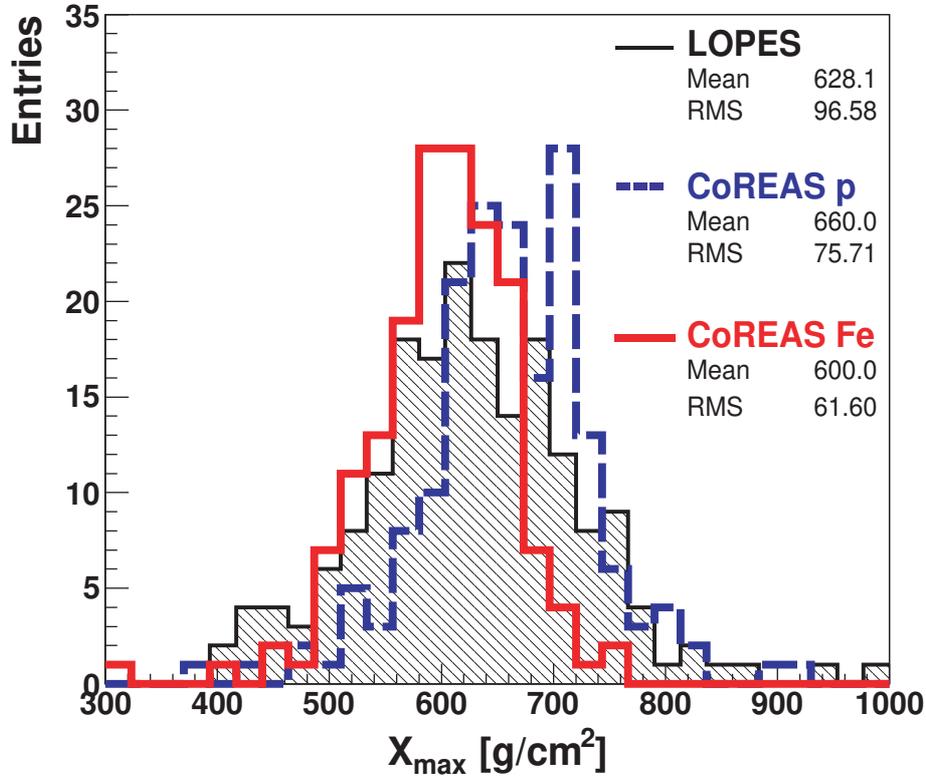}
  \caption{X$_{\mathrm{max}}$ reconstructed with the Slope Method for the LOPES measurements (black) and for CoREAS simulations (proton (dashed-blue) and iron (light-red)) of the same events, for the complete zenith angle range- 0-40 deg.}
   \label{xmax_histo}
\end{figure}

The X$_{\mathrm{max}}$ for LOPES is reconstructed by using eq.\ref{xmaxformula} and the $a$, $b$ and $c$ parameters derived with the CoREAS simulations. In fig.\ref{xmax_histo}, X$_{\mathrm{max, LOPES}}$ = 628.1 $\pm$ 96~g/cm$^{2}$, i.e. mean and standard deviation values. Compared to previous analysis  \cite{PalmieriArena,PalmieriThesis}, the expected improvement by using the most complete CoREAS simulations, is clearly achieved: The systematic shift shown in \cite{PalmieriArena} was definitively caused by the still existing discrepancy between the simulated and the measured LDF slope, due to the lack of refractive index effect in the previous analyses. These LOPES reconstructed X$_{\mathrm{max}}$ are compatible with the expectations from the cosmic ray nuclei and comparable with the CoREAS predicted values.

\section{Conclusion}
The Slope Method has been successfully applied also on the most recent CoREAS simulations of the LOPES events and, afterwards on the measurements themselves. Once again, this method reveals itself to be a powerful tool for both primary energy and X$_{\mathrm{max}}$ investigations with the radio data.

The specific distance from the shower axis d$_{\mathrm{0}}$ is re-confirmed with the LOPES measurements to be the best place for the primary energy analysis. The uncertainty for the energy reconstruction achieved is the combined energy resolution of both KASCADE and LOPES. The value is around 26\%, and it is comparable to the statistical uncertainty of the KASCADE experiment. One can conclude that the energy resolution of LOPES seams to be better than 26\%.

By using the updated CoREAS simulations, the systematic shift in the X$_{\mathrm{max}}$ reconstruction to smaller X$_{\mathrm{max}}$ values, typical of older analysis, is solved. The X$_{\mathrm{max}}$ obtained for the LOPES measurements are now comparable with expectations.\\ 
Moreover, an upper-limit precision of 96~g/cm$^{2}$ is found with the LOPES measurements, being the highest precision on X$_{\mathrm{max}}$ sensitivity to date achieved with the radio data.

Once more LOPES showed to be a powerful experiment to investigate properties of the radio emission from air showers, even though the high environmental noise limits its performance.

Higher precision is predicted by the Slope Method applied on pure simulations for both the energy ($\sim9\%$) and the X$_{\mathrm{max}}$ reconstruction ($\sim$50~g/cm$^{2}$). It may be achieved by experiments with a lower noise background.\\ 
This is the case for AERA (Auger Engineering Radio Array) \cite{AERAantennaPaper2012}, located at the Pierre Auger Observatory, and Tunka-Rex \cite{SchroederTunkarex_ICRC2013}, located in Siberia, which have the possibility to cross-check the reconstructed X$_{\mathrm{max}}$ values respectively with the fluorescence and the Cherenkov detectors.

\vspace*{0.5cm}
\footnotesize{{\bf Acknowledgment:}{LOPES and KASCADE-Grande have been supported by the German Federal Ministry of Education and Research. KASCADE-Grande is partly supported by the MIUR and INAF of Italy, the Polish Ministry of Science and Higher Education and by the Romanian Authority for Scientific Research UEFISCDI (PNII-IDEI grant 271/2011). This research has been supported by grant number VH-NG-413 of the Helmholtz Association.}}


\clearpage

\end{document}